\documentclass[sigconf]{acmart}
\usepackage{booktabs} 
\usepackage{soul}
\usepackage{multicol}
\usepackage{multirow}
\usepackage{balance}

\copyrightyear{2022}
\acmYear{2022}
\setcopyright{acmcopyright}
\acmConference[SAC '22]{The 37th ACM/SIGAPP Symposium on Applied Computing}{April 25--29, 2022}{Virtual Event}
\acmBooktitle{The 37th ACM/SIGAPP Symposium on Applied Computing (SAC '22), April 25--29, 2022, Virtual Event}
\acmPrice{15.00}
\acmDOI{10.1145/3477314.3507112}
\acmISBN{978-1-4503-8713-2/22/04}

\ccsdesc[500]{Computing methodologies~Artificial intelligence}
\ccsdesc[500]{Computing methodologies~Image segmentation}
\ccsdesc[300]{Computing methodologies~Supervised learning}
\ccsdesc[300]{Computing methodologies~Neural networks}
\ccsdesc[300]{Applied computing~Health informatics}

\begin{document}
\title[DAM-AL: Dilated Attention Mechanism with Attention Loss for 3D Infant Brain...]
{DAM-AL: Dilated Attention Mechanism with Attention Loss for 3D Infant Brain Image Segmentation}
  

\author{Dinh-Hieu Hoang}
\affiliation{%
  \institution{University of Science, Ho Chi Minh City, Vietnam}
}
\affiliation{%
  \institution{John von Neumann Institute, Ho Chi Minh City, Vietnam}
}
\affiliation{%
  \institution{Vietnam National University, Ho Chi Minh City, Vietnam}
}
\email{hieu.hoang2020@ict.jvn.edu.vn}

\author{Gia-Han Diep}
\affiliation{%
  \institution{University of Science, Ho Chi Minh City, Vietnam}
}
\affiliation{%
  \institution{John von Neumann Institute, Ho Chi Minh City, Vietnam}
}
\affiliation{%
  \institution{Vietnam National University, Ho Chi Minh City, Vietnam}
}
\email{han.diep@ict.jvn.edu.vn}

\author{Minh-Triet Tran}
\affiliation{%
  \institution{University of Science, Ho Chi Minh City, Vietnam}
}
\affiliation{%
  \institution{John von Neumann Institute, Ho Chi Minh City, Vietnam}
}
\affiliation{%
  \institution{Vietnam National University, Ho Chi Minh City, Vietnam}
}
\email{tmtriet@hcmus.edu.vn}

\author{Ngan T.H Le}
\affiliation{%
  \institution{University of Arkansas, Fayetteville, Arkansas USA}
}
\email{thile@uark.edu}

\begin{abstract}
While Magnetic Resonance Imaging (MRI) has played an essential role in infant brain analysis, segmenting MRI into a number of tissues such as gray matter (GM),  white matter (WM), and cerebrospinal fluid (CSF) is crucial and complex due to the extremely low intensity contrast between tissues at around 6-9 months of age as well as amplified noise, myelination, and incomplete volume. In this paper, we tackle those limitations by developing a new deep learning model, named DAM-AL, which contains two main contributions, i.e., dilated attention mechanism and hard-case attention loss. Our DAM-AL network is designed with skip block layers and atrous block convolution. It contains both channel-wise attention at high-level context features and spatial attention at low-level spatial structural features. Our attention loss consists of two terms corresponding to region information and hard samples attention. Our proposed DAM-AL has been evaluated on the infant brain iSeg 2017 dataset and the experiments have been conducted on both validation and testing sets. We have benchmarked DAM-AL on Dice coefficient and ASD metrics and compared it with state-of-the-art methods. Code is available at:
\href{https://github.com/DinhHieuHoang/DAM-CA-InfantBrain}{https://github.com/DinhHieuHoang/DAM-CA-InfantBrain}
\end{abstract}

\keywords{Attention Loss, Channel-wise Attention, Spatial Attention, Infant Brain, Segmentation}

\maketitle

\section{Introduction}

Accurate infant brain MRI segmentation is a crucial modern technique in recognizing normal and abnormal early brain development \cite{mittal2019deep}. For instance, the report \cite{hazlett2011early} shows that brain overgrowth is associated with an increase in the cortical surface area before two years of age in autistic children. One of the most critical procedures to measure infant brain development and identify biomarkers is accurate segmentation of MRI into different tissue areas, i.e., white matter (WM), gray matter (GM), and cerebrospinal fluid (CSF) \cite{knickmeyer2008structural}. Infant brain segmentation is considered to be more challenging than adult brain segmentation due to tissue contrast reduction, amplified noise, myelination, and incomplete volume \cite{wang2019quantitative}. Furthermore, the intensity distributions of GM and WM have larger overlapping; thus it is difficult for manual annotation. Hence annotated data is highly limited. 

Deep learning, i.e. Convolutional neural networks (CNNs) have obtained great achievement in computer vision including in medical imaging tasks. To address those problems, several approaches leveraged deep learning have been proposed to improve infant brain segmentation accuracy \cite{cciccek20163d, jegou2017one, nie20183, le2018deep, bui2019skip, chen2018voxresnet, le2020offset, le2021multi, le2021narrow, le2021offset, yamazaki2021invertible, nguyen20213d}. We can generally divide the existing works into two categories: the first category focuses on the network architecture designs, whereas the second category targets proposing loss functions. In the first category, \cite{zhang2015deep, moeskops2016automatic} first proposed using 2D CNNs to segment isointense-phase brain images. However, these 2D CNNs are time-consuming since they process each slice independently and fail to capture the spatial contextual information present in the volumetric data. Thus, a 3D network structure has been later developed and outperformed most 2D network architectures. For instance, \cite{cciccek20163d} replaced all 2D operations of the U-Net architecture \cite{2Dunet} with 3D counterparts for volumetric biomedical image segmentation without increasing the number of parameters. Later, \cite{jegou2017one} proposed a tiramisu network as a fully convolutional densenets; \cite{nie20183} proposed 3D FCN for multimodal isointense infant brain segmentation; \cite{bui2019skip} proposed skip-connected 3D DenseNet; \cite{dolz2018hyperdense} introduced HyperDenseNet having complex dense connections between paths of different modalities. In the second category, 3D Unet \cite{3Dunet} is used as a backbone network, and various proposed loss functions have been evaluated against existing loss functions. Notably, \cite{le2020offset} proposed offset curves loss and \cite{le2021narrow} proposed NB-AC loss for medical image segmentation, and they both mainly target imbalanced class problem. Beyond CNNs, Capnets\cite{sabour2017dynamic} has been improved to 3D volumetric data to address infant brain segmentation \cite{nguyen20213d}. Some more advanced deep-learning-based methods have also been reported in the iSeg-2017 challenge review \cite{wang2019benchmark}.

\begin{figure*}[t!]
    \centering
    \includegraphics[width=0.98\textwidth]{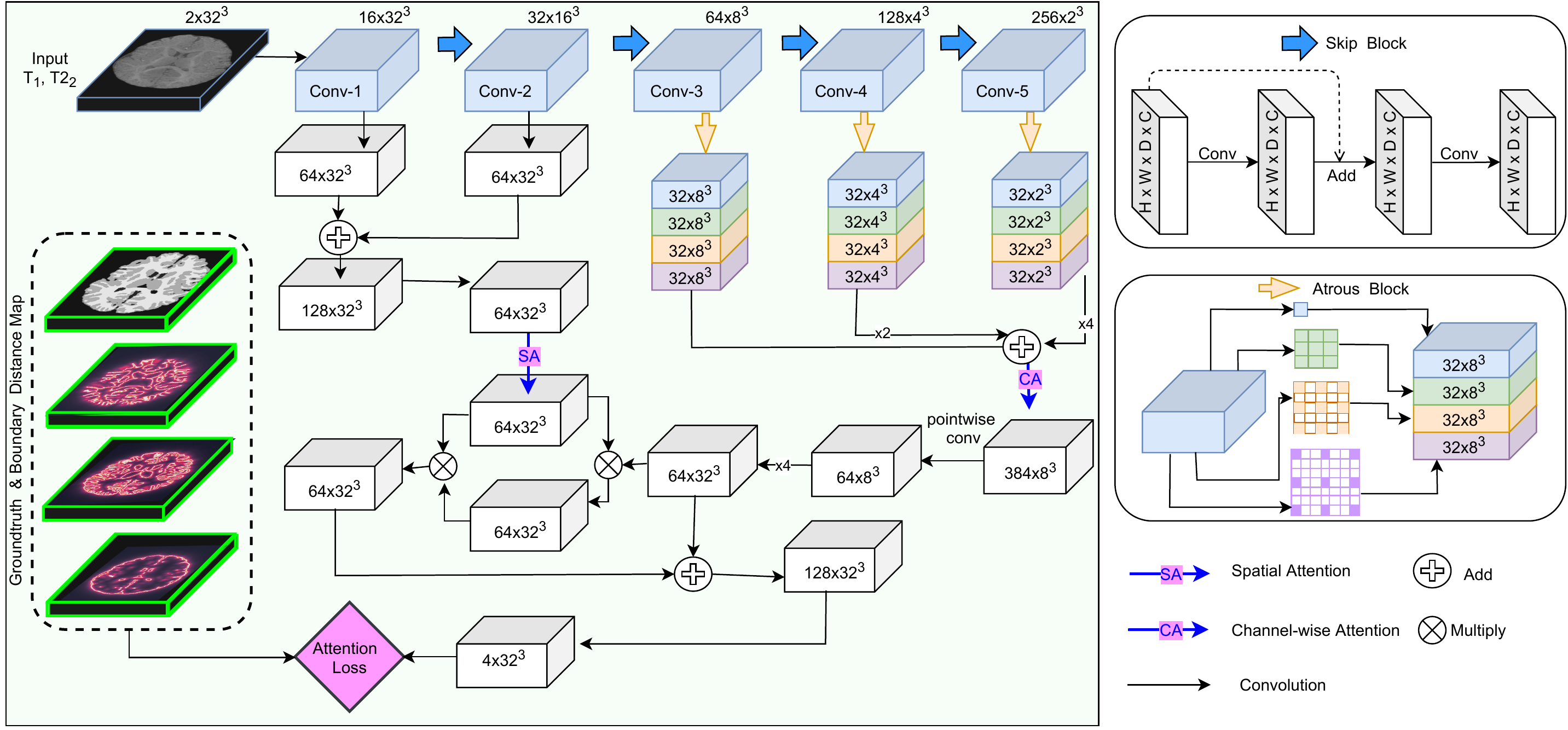}
    \caption{The overall architecture of our proposed DAM-AL. The encoder path is leveraged by Unet contracting path with skip block connections. The low-level layers i.e. Conv-1, Conv-2 are modeled to form low-level feature through spatial attention (SA). The high-level layers i.e. Conv-3, Conv-4, Conv-5 are learnt to generate high-level feature by channel-wise attention (CA). Atrous block is designed to weight high-level layers at multi-scale as well as to extract richer context-aware information.}
    \label{fig:flowchart}
\end{figure*}

To the best of our knowledge, all existing architectures used in volumetric brain tissue segmentation are based on Unet architecture which contains two paths: encoder path downsamples feature maps to capture the contextual information and decoder path upscales the downsampled feature maps for localization. In those networks, skip-connections is utilized to facilitate information flow from the encoder path to the decoder path. However, those networks treat all data points i.e. voxels equally and there is no mechanism to pay attention on hard data points which are easily misclassified. Infant brain data is captured in low contrast and weak surface, thus segmenting the areas around surface considered as hard-case samples is very challenging. Furthermore, most DL-based segmentation networks have made use of common loss functions, e.g., CE, Dice, Focal. These losses are based on summations over the segmentation regions and are restricted to pixel-wise settings. Not only pixel-wise sensitivity, but these losses are also unfavorable to low contrast which is a big challenge in infant brain segmentation. Furthermore, these losses are working on higher level features of region information and none of them is intentionally designed for lower level features such as edge/surface which play an important role in medical imaging.

In this work, we first present an effective network architecture which is based on UNet \cite{cciccek20163d} and influenced by spatial and channel-wise attention network \cite{chen2017sca}. Our DAM-AL network contains skip blocks and atrous blocks layers to capture the rich context information at both high-level and low-level features. We then introduce our hard-case attention loss, defined as a surface distance map weighted by hard cases estimation function. Our contribution is summarized as follows:
\begin{itemize}
    \item Introduce a dilated attention network consists of skip block connection and atrous block layers
    \item Present spatial attention and channel-wise attention to enrich the high-level context features and low-level spatial structural features.
    \item Introduce a hard-case attention loss which addresses the low contrast and weak surface problem in infant brain image segmentation.
    \item The proposed DAM-AL outperforms other state-of-the-art methods on various metrics.
\end{itemize}

\section{Proposed Method}
\subsection{Dilated Attention Network}
Attention mechanisms have been successfully applied in various computer vision tasks such as object recognition \cite{mnih2014recurrent}, image captioning \cite{chen2017sca}, action detection \cite{vo2021agent}. In this paper, we leverage SCA-CNN network \cite{chen2017sca} and propose a dilated attention network that contains channel-wise attention (CA) mechanism to capture context at high level feature and spatial attention (SA) mechanism to capture structure at low level feature. The entire DAM-CA is illustrated in Fig.\ref{fig:flowchart}.

\subsubsection{Multiscale feature extraction}
Our feature extraction is based on Unet encoder which contains five layers with skip-block connection. The first two layers (low-level layers) (conv-1, conv-2) conduct a low level feature while the last three layers (high-level layers) (conv-3, conv-4, conv-5) produce a high level feature.

\subsubsection{Spatial attention (SA) at the low-level layers:}
\begin{figure*}[!h]
    \centering
    \includegraphics[width=0.85\textwidth]{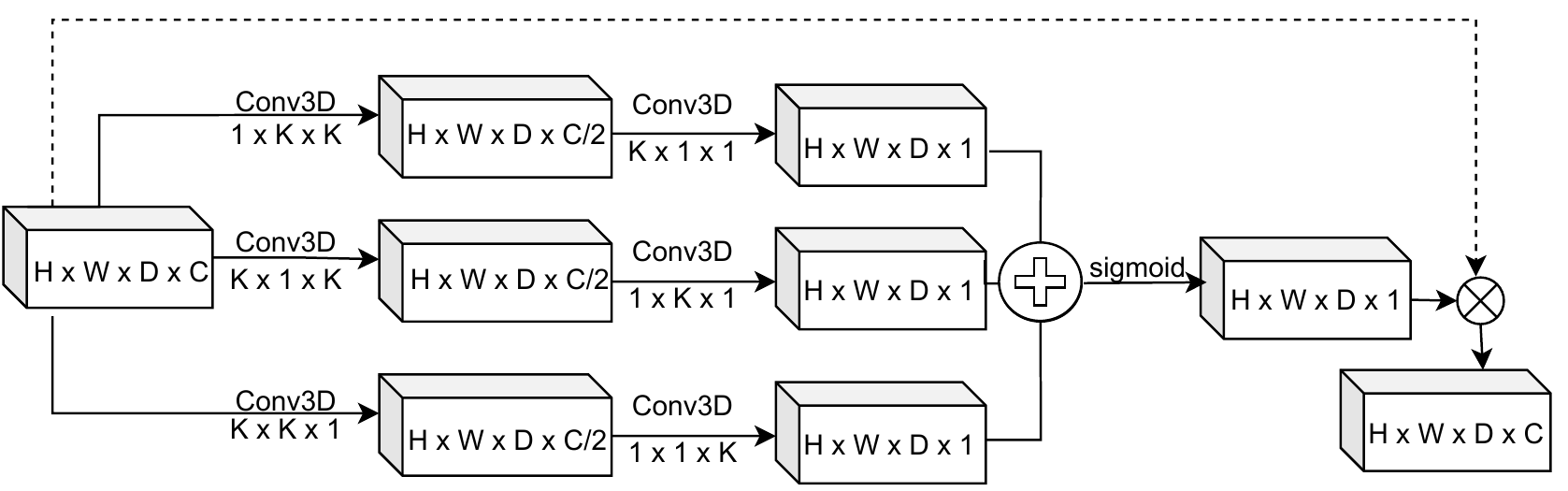}
    \caption{The illustration of spatial attention (SA).}
    \label{fig:SA}
\end{figure*}

In medical image segmentation, boundaries between objects play an important role; thus, we want to obtain as detailed as possible the boundaries. Instead of considering all spatial positions equally, we adopt SA \cite{sca_gate} to focus more on the foreground regions, especially surface areas. Let denote $F_l\in R^{H\times W\times D \times C}$ as feature at a low-level layer. Each voxel in spatial is presented as $(x,y,z)$ in the spatial coordinate. To increase receptive field and extract better global information but not increase parameters, we apply six convolution layers. The kernels of these layers are $1\times K \times K$, $K \times 1 \times 1$, $K \times 1 \times K$, $1 \times K \times 1$, $K \times K \times 1$, and $1 \times 1 \times K$. Fig.\ref{fig:SA} illustrates SA applying into the low-level layers to produce low-level features. The final low-level feature is obtained by weighting $F_l$ with SA feature $F_{sa}$. The process can be presented as follows:
\begin{equation}
\begin{split}
    F_1 & = conv2(conv1(F_l)); 
    F_2  = conv4(conv3(F_l)); \\
    F_3 & = conv6(conv5(F_l)); 
    F_{sa}  = sigmoid(F_1 + F_2 + F_3); \\
    F_L & = F_{sa}F_l.
\end{split}    
\end{equation}
where $conv1, conv2, conv3, conv4, conv5, conv6$ are corresponding to different kernels. 

\subsubsection{Channel-wise attention (CA) at the high-level layers:} 
\begin{figure*}[!h]
    \centering
    \includegraphics[width=.65\textwidth]{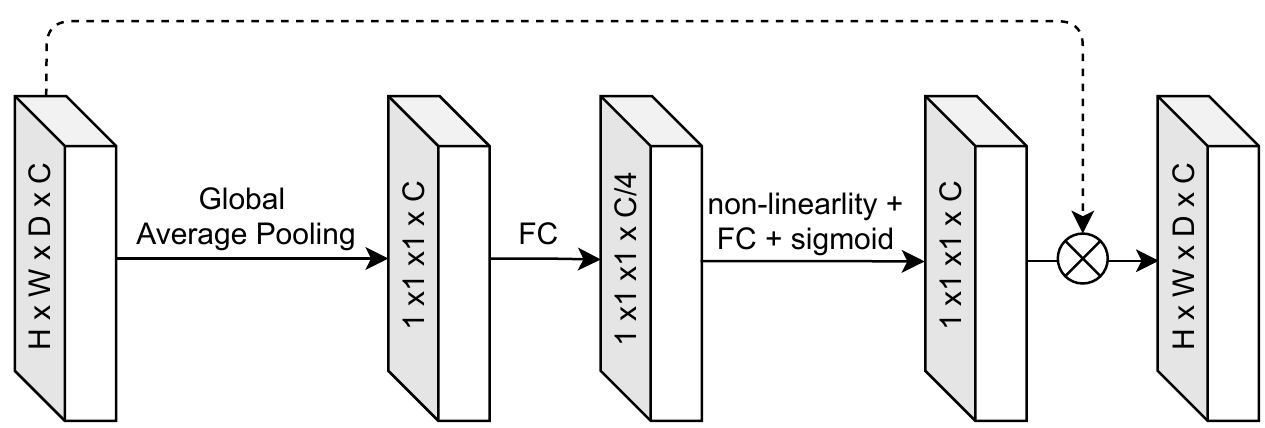}
    \caption{The illustration of channel-wise attention (CA).}
    \label{fig:CA}
\end{figure*}

Because different channels of features in CNNs present different semantics, we apply channel-wise attention (CA) \cite{sca_gate} to weighted multi-scale at high-level layers. To capture richer structure information with longer range dependence, we adapt atrous convolutions with different dilation rates at the high-level layers. Notably that the CA will assign larger weights to channels that present high response to surface and region-of-interest i.e. WM, GM, CSF. In our case, the dilation rates set to 1, 2, 3, 4. 
The feature maps from different atrous convolutional layers are then combined by concatenation. Let denote $F_h\in R^{H\times W\times D \times C}$ as feature at a high-level layer, the high-level feature $F_H$ is obtained by weighting $F_h$ with CA feature $F_{ca}$. The procedure is illustrated as in Fig.\ref{fig:CA} and formulated as follows:
\begin{equation}
    \begin{split}
        F_{ca} & = sigmoid(FC(\delta(FC(Pool(F_h))))); \\
        F_H & = F_{ca}F_h.
    \end{split}
\end{equation}
where $FC$, $\delta$ are fully connected layer and non-linearity function (reLU). 

\subsection{Hard-case Attention Loss}

To train a Deep Neural Network (DNN), the loss function, known as cost function, plays a significant role. 
The loss function is to measure the average (expected) divergence between the output of the network ($P$) and the ground truth ($T$) being approximated over the entire domain of the input, sized $W \times H \times D$. 
We denote $i$ as index of each voxel in an volumetric medical image spatial space $N = W \times H \times D$. 
The label of each class is written as $c$ in $C$ classes. 

Herein, we first briefly review and analyze some common loss functions, introduce hard-case samples estimation, and finally present our attention loss function.

\subsubsection{Existing loss functions} Cross Entropy (CE), Dice loss, and Focal loss are common loss functions in image segmentation while offset loss, boundary loss are the state-of-the-art loss functions that address the problems of imbalanced data and weak surface in medical image segmentation. 

\begin{itemize}
 \item {Cross Entropy (CE) Loss:} it was proposed by \cite{murphy2012machine}, and is a widely used pixel-wise distance to evaluate the performance of the classification or segmentation model. It is defined as $\mathcal{L}_{CE} = -\frac{1}{N}\sum_{i=1}^{N}{[T_i \ln(P_i) + (1 - T_i) \ln(1-P_i)] }$. However, for unbalanced data, it typically results in unstable training results and leads to decision boundaries biased towards the majority classes. To deal with the imbalanced-data problem, two variants of the standard CE loss, Weighted CE (WCE) loss and Balanced CE (BCE) loss, are proposed to assign weights to the different classes.
\item {Dice loss:} it was proposed 
by \cite{Milletari_2016} and defined as $\mathcal{L}_{Dice} = 1 - 2\frac{\sum_i^N{T_i P_i}}{\sum_i^N{T_i+P_i}} = 2\frac{T \cap P}{T \cup P}$. Despite the Dice loss improvements over the CE loss, Dice loss may undergo difficulties when dealing with very small structures and weak object boundary, as misclassification of a few pixels can lead to a large decrease of the coefficient.
\item {Focal Loss:} it was proposed by \cite{Lin_2018} to balance between easy and hard samples as $\mathcal{L}_{Focal} = -\frac{\alpha_i}{N}\sum_{i=1}^{N} \big((1-P_i)^\gamma T_i \ln(P_i) +P_i^\gamma ( 1 - T_i) \ln(1- P_i)\big)$. In Focal loss, the loss for confidently correctly classified labels is scaled down, so that the network focuses more on incorrect and low confidence labels than on increasing its confidence in the already correct labels.
 \item {Boundary Loss:}
Recently, Kervadec, et al. \cite{Kervadec_2021} and Le et al. \cite{le2021offset, le2021narrow} proposed boundary loss, offset curve loss (OsC) loss, and NB-AC loss to address both imbalanced data and weak boundary problems. For instance, OsC loss focuses on narrow band around the boundary. Boundary loss \cite{Kervadec_2021} makes use of a distance map in which the weights of voxels far from the boundary with greater than the nearer ones, thus the model is learnt to reduce the error on the midst of the regions rather than the boundary. Despite its plausible intuition and competitive performance in practice, it does not help much in dealing with low contrast and weak boundary. 
\end{itemize}

\subsubsection{Attention Loss} 

Our attention loss is leveraged by \cite{Kervadec_2021, le2020offset, le2021narrow} to pay more attention to hard-case examples or easily misclassified examples i.e. regions which are poorly segmented. In this section, we first show how to estimate hard-case examples and then present our proposed hard-case attention loss.

\textbf{Hard-case examples estimation:}

Infant brain MRI is shown in low contrast, weak boundary/surface; thus hard-case examples are more often presented at the surface regions. We also further observe that the model quickly learns and produces high accuracy on easy-case voxels (inside the object). Still, it is time-consuming and performs poorly on hard-case voxels at the surface. Therefore, it is reasonable to force the model to concentrate on the hard-case voxels while reducing the effect of easy-case voxels in updating the model's parameters. When it comes to Focal loss \cite{Lin_2018}, the authors introduce a new term and integrate it into CE loss so that the model turns its attention from the subjects which are correctly classified with high confidence to the incorrect ones. The motivation of this term is to increase the gradient magnitude of the loss function of hard-case examples and decrease that of the easy-case ones so that the total gradient related to low confident and incorrect labels dominates the total gradient yielded by the correct ones. To make this idea more solid, we compare CE loss and Focal loss. To simply the equation, let consider one voxel at spatial coordinate $i = (x,y,z), x = 1, .., H, y = 1, .., W, z = 1, .., D$ with ground truth label $T_i$ and predicted label $P_i$. The CE loss and Focal loss at $i$ are as follows:
\begin{equation}
\begin{split}
    \mathcal{L}_{CE} & = -\big[T_i \ln(P_i) + (1 - T_i) \ln(1-P_i)\big].\\
\mathcal{L}_{Focal} & = -\big[(1-P_i)^\gamma  T_i \ln(P_i)  +P_i^\gamma ( 1 - T_i) \ln(1- P_i)\big].
\end{split}
\end{equation}

Without any loss of generality, we consider the case where $T_i=1$ and corresponding losses are simplified into:

\begin{equation}
\begin{split}
\mathcal{L}_{CE} & = -\ln(P_i). \\
\mathcal{L}_{Focal} & = -(1-P_i)^\gamma\ln(P_i).
\end{split}
\end{equation}

The derivatives of divergence functions w.r.t $P_i$ are as follows:

\begin{equation}
\begin{split}
\frac{\partial\mathcal{L}_{CE}}{\partial P_i} & = -\frac{1}{P_i}. \\
\frac{\partial\mathcal{L}_{Focal}}{\partial P_i} & = -\frac{(1-P_i)^\gamma}{P_i} + \gamma(1-P_i)^{\gamma-1}\ln(P_i).
\end{split}
\end{equation}

For $\gamma=2$ which is the optimal value in \cite{Lin_2018}, the absolute value of the gradient of $\mathcal{L}_{Focal}$ is greater than that of $\mathcal{L}_{CE}$ when $P$ is smaller than approximately 0.298. Therefore, the model focuses on the hard examples whose predicted probability of belonging to true classes is smaller than $29.8\%$.

\textbf{Hard-case attention loss:}
In this section, we first define a surface attention weight map $W^c$ for each class $c\in C$. Let $T^{c}$ denote as surface of class $c$. Each element $W^c_i$ at spatial coordinate $i$ in the surface attention weight map $W^c$ is defined as: 

\begin{equation}
    \mathcal{W}^c_i=\frac{1}{min_{j\in T^c}d(i,  j) + 1}.
\end{equation}
where $d$ is defined as ${\mathcal{L}_2}$ and $\mathcal{W}^c_i$ measures the distance between surface and voxels in the spatial domain. Unlike existing boundary loss functions, the voxels near the surface receive more weight than those far from the surface in our proposed weight map. Furthermore, to maintain the attention on the surface without introducing noise into the model, we incorporate Dice loss, which helps noise-free regions inside surface. 
\begin{figure}[!h]
    \centering
    \includegraphics[width=0.48\textwidth]{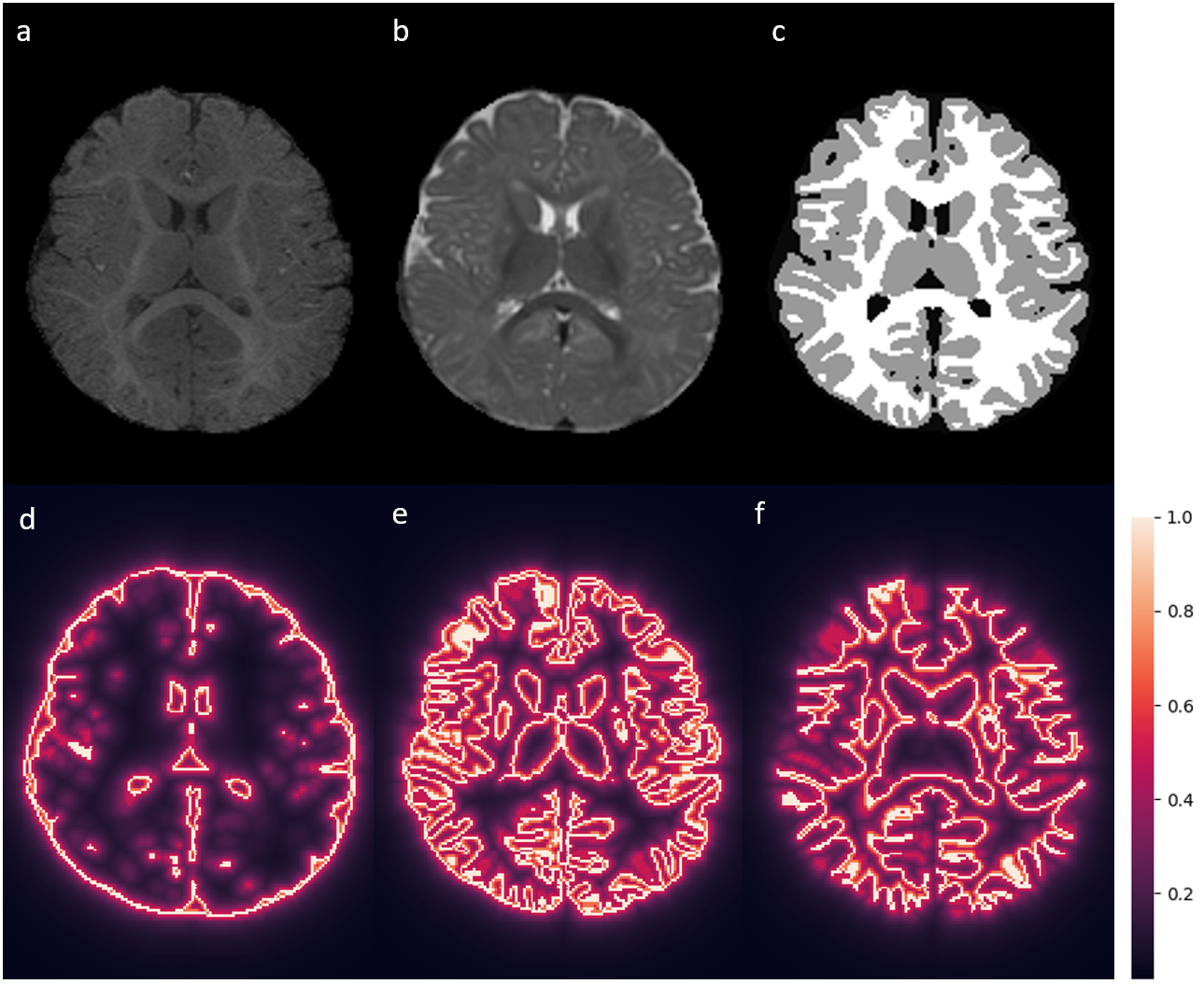}
    \caption{(a) and (b) are T1- and T2-weighted brain MRI scans of the subject 9 in iSeg2017 dataset. (c) is its segmentation ground truth, and (d), (e), and (f) are the corresponding surface attention weight map for cerebrospinal fluid (CSF), gray matter (GM) and white matter (WM).}
    \label{fig:weight_map_illustration}
\end{figure}

\begin{table*}[!h]
    \centering
    \caption{Comparison on iSeg-2017 dataset with Experiment Setting 1: Train on 9 subjects and test on Subject \#9. The best is shown in bold and the second best is shown in \underline{underline}.}
    \label{table:iseg}
    \begin{tabular}{l|l|llll}
        \toprule
        \multicolumn{1}{c}{\multirow{2}{*}{Method}} & Year &  \multicolumn{4}{c}{DSC $\uparrow$}                                        \\ \cmidrule(l){2-5} 
       &  & WM & GM & CSF & Average \\
        3D-UNet \cite{cciccek20163d}  & 2016& 89.83  & 90.55 & 94.39 & 91.59  \\
        DenseVoxNet \cite{jegou2017one} & 2017 & 85.46 & 88.51 & 91.26 & 89.24  \\
        
        VoxResNet \cite{chen2018voxresnet} & 2018& 89.87 & 90.64 & 94.28 & 91.60    \\
        CC-3D-FCN \cite{nie20183}  & 2018 & 89.19 & 90.74 & 92.40 & 90.79 \\
        SegCaps \cite{lalonde2018capsules}  & 2018 & 82.80 & 84.19 & 90.19 & 85.73\\
        3D-SkipDenseSeg \cite{bui2019skip}& 2019& \underline{91.30} & 91.61 & 94.74      & 92.55 \\ 
        SemiDenseNet \cite{SemiDenseNet} & 2020& 90.50 & \textbf{92.05} & \textbf{95.80} & \underline{92.77} \\ 
        3D-UCaps \cite{nguyen20213d} & 2021& 90.95 & 91.34 & 94.21 & 92.17 \\
        MSCD-UNet \cite{long2021learning} & 2021 & 90.47 & 92.17 & \underline{95.60} & 92.74 \\
        \midrule
        Our proposed  & & \textbf{91.36} & \underline{91.92} & 95.06 & \textbf{92.78} \\
        \bottomrule
    \end{tabular}
    \label{tab:valset}
\end{table*}

\begin{figure*}[!h]
    \centering
    \includegraphics[width=0.95\textwidth]{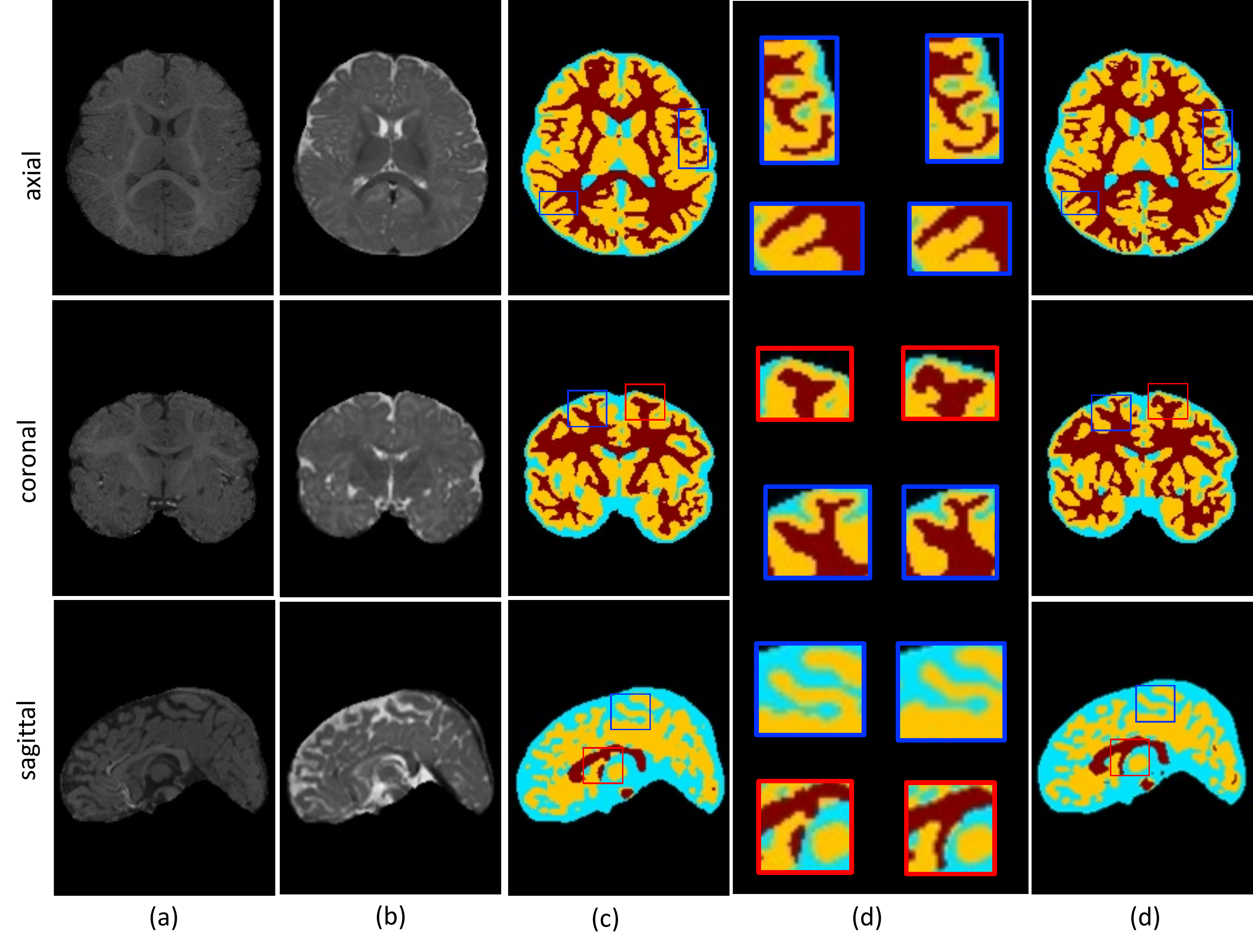}
    \caption{Result of Experiment Setting 1. (a) and (b) are T1- and T2-weighted brain MRI scans of the subject \#9 in different views. (c): predicted segmentation results by DAM-AL; (d) Enlarged view of some random regions between (c) and (e). Blue boxes indicate some spots produced correctly by DAM-AL. Some regions where DAL-AL yielded incorrect segmentations are outlined in red ; (e): ground truth. Top-down: visualize in different planes: axial, coronal, and  sagittal. }
    \label{fig:result_experiment_1}
\end{figure*}

\begin{table*}[!h]
    \centering
    \caption{Comparison on iSeg-2017 dataset with Experiment Setting 2: Train on 10 subjects and test on 13 Subjects. The best is shown in bold and the second best is shown in \underline{underline}.}
    \label{table:iseg}
    \begin{tabular}{l|l|llll|llll}
        \toprule
        \multicolumn{1}{c}{\multirow{2}{*}{Method}} & Year & \multicolumn{4}{c}{DSC $\uparrow$}  & \multicolumn{4}{c}{ASD (mm)$\downarrow$}                                        \\ \cmidrule(l){2-9} 
       &  & WM & GM & CSF & Average &  WM & GM & CSF & Average \\
        HyberDense  \cite{dolz2018hyperdense} & 2018& 90.1 & 92.0 & 95.6 & 92.57 & 0.38 & 0.32 & \underline{0.12} & 0.27 \\
        FC-DenseNet  \cite{hashemi2019exclusive} & 2019& 90.7 & 92.6 & \textbf{96.0} & 93.1 & 0.36 & \underline{0.31} & \textbf{0.11} & \underline{0.26} \\
        D-SkipDenseSeg  \cite{bui2019skip} & 2019& 90.3 & 92.2 & 95.7 & 92.73 & 0.38 & 0.32 & \underline{0.12} & 0.27\\
        H-DenseNet  \cite{qamar2019multi} & 2019& 90.0 & 92.0 & \textbf{96.0} & 92.67 & 0.36 & \underline{0.31} & \textbf{0.11} & \underline{0.26} \\
        FC-Semi-DenseNet1 \cite{SemiDenseNet}& 2020 & 90.0 & 92.0  & \textbf{96.0} & 92.67 & 0.38 & 0.35 & 0.14 & 0.29 \\
        FC-Semi-DenseNet2  \cite{SemiDenseNet}& 2020 & 90.0 & 92.0  & \textbf{96.0} & 92.67 & 0.41 & 0.34 & \underline{0.12} & 0.29 \\
        V-3D-UNet  \cite{qamar2020variant} & 2020& 91.00 & 92.00 & \textbf{96.00} & \underline{93.00} & 0.37 & \underline{0.31} & 0.13 & 0.27 \\
        Non-local U-Net \cite{wang2020non} & 2020 & 91.03 & \underline{92.45} & 95.30 & 92.29 & 0.40 & 0.37  & 0.14 & 0.30\\
        HyperFusionNet \cite{duan2021multi} & 2021& \underline{90.20} & 87.80 & 93.60 & 90.53 & -- & -- & -- \\
        APRNet\cite{zhuang2021aprnet} & 2021 & 91.10 & 92.40 & 95.50 & \underline{93.00}& \underline{0.35} & 0.32 & 0.12 & \underline{0.26} \\
       \midrule
        Our proposed  & & \textbf{92.60} & \textbf{93.49} & \underline{95.68} & \textbf{93.92} & \textbf{0.28} & \textbf{0.25} & \textbf{0.11} & \textbf{0.21} \\  
        \bottomrule
    \end{tabular}
    \label{tab:testset}
\end{table*}

\begin{figure*}[!h]
    \centering
    \includegraphics[width=0.9\textwidth]{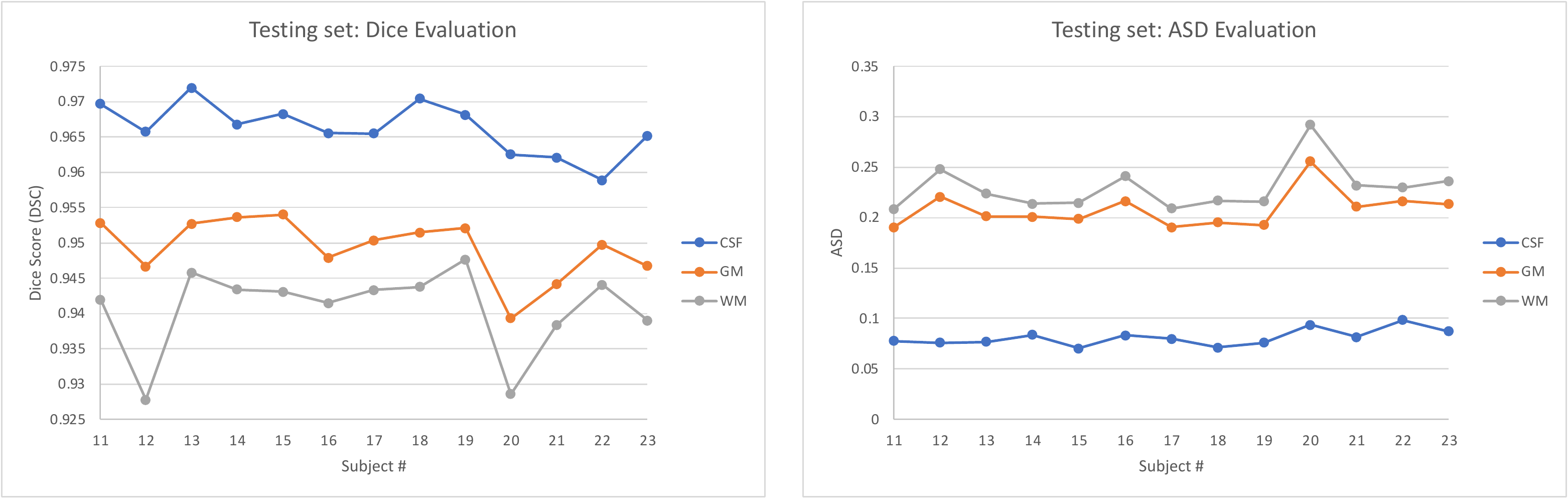}
    \caption{Performance of the proposed DAM-AL on the subjects of iSeg-2017 datasets. left: DSC, right: ASD.}
    \label{fig:testing_performance}
\end{figure*}

\begin{figure*}[!h]
    \centering
    \includegraphics[width=0.95\textwidth]{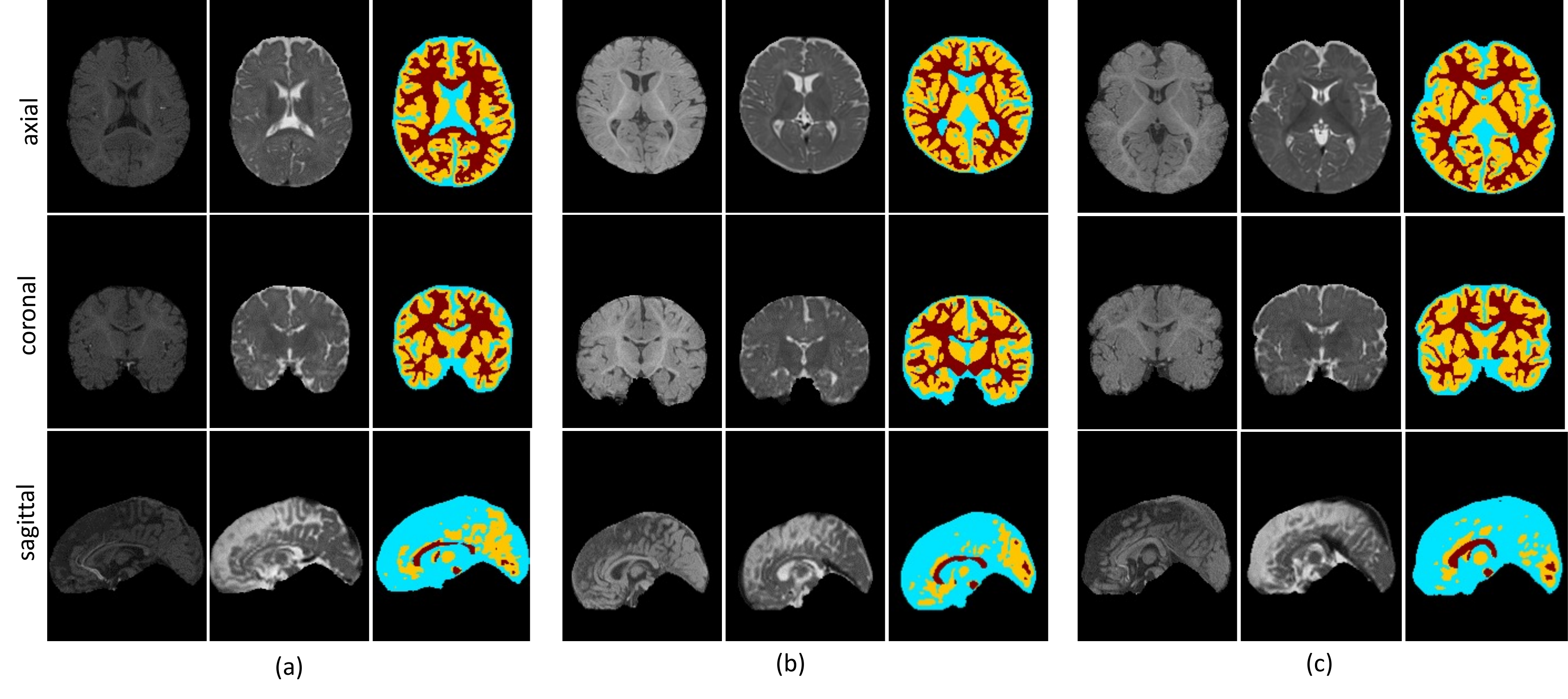}
    \caption{Result of Experiment Setting 2 with the first three subjects, i.e. \#11 (a) , \#12 (b) , \#13 (c). Top-down: visualize in different planes: axial, coronal, and sagittal. For each subject, from left-right: T1-brain MRI, T2-brain MRI, predicted segmentation results conducted by DAM-AL. }
    \label{fig:result_experiment_2}
\end{figure*}

Figure \ref{fig:weight_map_illustration} illustrates the low contrast, weak boundary problem in brain infant MRI segmentation. The most challenging is how to decide the voxels on the surfaces where they belong to the left class or right class. The weight maps (d, e, f) corresponding to CSF, GM, and WM surfaces show the significance of voxels. The voxels far from surfaces have less impact than those which are close to the surface. 

Incorporate hard-case estimation into weight map, our attention loss is defined as follows:

\begin{equation}
\mathcal{L}_{attention}= \begin{cases} \sum_{i\in P, c\in C}W^{c}_i(P^c_i)^2 & \quad \text{if }P_{i}\neq c\\
\sum_{i\in P, c\in C}W^{c}_i(1-P^c_i)^2& \quad \text{otherwise} \end{cases}
\end{equation}

The derivative of our $\mathcal{L}_{attention}$ w.r.t $P_i$ corresponding to a particular class $c$ is:
\begin{equation}
    \frac{\partial f_{Attention}}{\partial P_i}=-2(1-P_i).
\end{equation}

In our attention loss function, the voxels which are wrongly classified (predicted probability of being in the same class as ground truth smaller than $50\%$) will be strongly attended. Furthermore, our proposed loss will emphasize the true class of each voxel as well as strengthen the confidence of the model prediction without having too much negative effect on rare classes.

\section{Experimental Results}
\subsection{Data and metrics}

\noindent 
\textbf{Dataset:} The iSeg17 dataset \cite{wang2019benchmark} consists of 10 subjects with ground-truth labels for training and 13 subjects without ground-truth labels for testing. Each subject includes T1 and T2 images with a size of $144 \times 192 \times 256$, and an image resolution of $1 \times 1 \times 1$  $\text{mm}^3$. In iSeg, there are three classes:  white matter (WM), gray matter (GM), and cerebrospinal fluid (CSF).

\noindent 
\textbf{Metrics:}
For quantitative assessment of the segmentation, the proposed model is evaluated on different metrics, e.g. Dice score (DSC), and average surface distance (ASD). 

The DSC measure is defined as:
\begin{equation}
    DSC \frac{2|T \cap P|}{|T| + |P|}.
\end{equation}
where $T$ and $P$ are corresponding to groundtruth and predicted segmentation result. A higher value of DSC means better segmentation accuracy.

The ASD is utilized to measure the segmentation boundary distance and defined as:
\begin{equation}
\small
    ASD(T, P) = \frac{1}{2}\left(\frac{\sum_{V_i\in S_P}{min_{V_j \in S_T}d(V_i, V_j)}}{\sum_{V_i\in S_P}{1}} + \frac{\sum_{V_j\in S_T}{min_{V_i \in S_G}d(V_i, V_j)}}{\sum_{V_j\in S_T}{1}}\right).
\end{equation}
where $S_T$ and $S_P$ are the surface of the ground truth and predicted segmentation result. $d(V_i, V_j)$ is the Euclidean distance from a vertex $V_i$ and $V_j$. A smaller ASD is the better the result.

\subsection{Experiment Setting}
The input is defined as $N \times C \times H \times W \times D$, where $N$ is the batch size, $C$ is the number of input modalities, and $H, W, D$ are height, width, and depth of the volume patch in the sagittal, coronal, and axial planes. We choose the  input as $8 \times 2 \times 32 \times 32 \times 32$. We implemented our network using PyTorch 1.3.0, and our model is trained until convergence by using the SGD optimizer accompanied by the warm restarts technique. More precisely, the training phase consists of 50 200-epoch periods, in which the learning rate is set to $0.01$ and reduces by ten times every 40 epochs. As aforementioned, medical images are widely considered as difficult subjects for CNN models to learn meaningful and useful characteristics. In other words, their special domain is far from that of natural images on which common deep learning models and modules are designed. Intuitively, in the training phase, the larger the searched space is, the higher probability the model learns special features, which help to solve the problem efficiently. By repeatedly restarting the learning rate schedule, the model is encouraged to explore a larger space every time the learning rate is set to maximum. After going to a new area in the search space, the learning rate rapidly drops to find the local optimal parameters in this area. Our DAM-AL makes use of instance normalization \cite{instance_normalization} and Leaky reLU. The experiments are conducted using an Intel CPU and RTX GPU with two settings. 

\begin{itemize}
    \item Experiment Setting 1: We follow 3D-SkipDenseSeg \cite{bui2019skip} to have the training set of 9 subjects and testing set of subject \#9. 
    \item Experiment Setting 2: We train on 10 annotated subjects and test on  13 unlabeled subjects (Subject \#11 - Subject \#23). 
\end{itemize}

\subsection{Performance and Comparison}
The evaluations on iSeg-2017 are given in Table\ref{tab:valset} and Table\ref{tab:testset} corresponding to two experiment settings. Regarding Experiment Setting 1, our DAM-AL obtains the best performance on WM and the second-best on GM and CSF compared with the existing state-of-the-art approaches. Overall, our  DAM-AL outperforms other methods on DSC in Experiment Setting 1. Fig.\ref{fig:result_experiment_1} visualizes the performance of DAM-AL on Subject \#9 in the coronal, axial, and sagittal planes. On Subject \#9, we randomly crop some patches and compare the performance between the predicted segmentation and ground truth at an enlarged view. In most of cases, the hard-case examples on surface are predicted correctly. Regarding Experiment Setting 2, which was evaluated on testing set, our DAM-AL obtains the best score on both DSC and ASD metrics with considerable gaps compared to the second-best performance, i.e. average DSC gains 0.92\% compared with the second-best \cite{qamar2020variant} and average ASD reduces 0.05mm compared with the second-best \cite{hashemi2019exclusive, qamar2019multi, zhuang2021aprnet}. Performance on DSC and ASD of individual subject is provided in Fig. \ref{fig:testing_performance}. Fig.\ref{fig:result_experiment_2} visualizes the performance of DAM-AL on Subject \#11.


\subsection*{Conclusion}
In this work, we introduced DAM-AL, a dilated attention mechanism with attention loss on hard-case examples for medical image segmentation. Our proposed DAL-AL contains spatial attention at high-level feature and channel-wise attention at low-level feature. We tested our framework on the problem of infant brain segmentation and showed that our DAM-AL is effective, robust, and more accurate than existing segmentation methods.

Therefore, exploring spatial and channel-wise attention together with hard-case attention loss is a promising approach to medical image analysis. Future investigations might include other medical image datasets on different modalities, e.g., MRBrainS, Brats, Pancreas, Hippocampus, etc.

\balance

\section*{Acknowledgment}
This material is based upon work supported by the National Science Foundation under Award No. OIA-1946391,  
partially funded by Gia Lam Urban Development and Investment Company Limited, Vingroup and supported by Vingroup Innovation Foundation (VINIF) under project code VINIF.2019.DA19.
Dinh-Hieu Hoang and Gia-Han Diep were funded by Vingroup Joint Stock Company and supported by the Domestic Master/ PhD Scholarship Programme of Vingroup Innovation Foundation (VINIF), Vingroup Big Data Institute (VINBIGDATA), code VINIF.2020.ThS.JVN.02 and VINIF.2020.ThS.JVN.04, respectively.

\bibliographystyle{ACM-Reference-Format}
\bibliography{sample-bibliography} 

\end{document}